\documentclass[]{aastex631}
\shortauthors{Li et al.}
\graphicspath{{./}{figures/}}
\usepackage{courier}
\usepackage{longtable}
\begin{document}

\title{Five New Heartbeat Star Systems with Tidally Excited Oscillations Discovered Based on TESS Data}

\author[0000-0002-8564-8193]{Min-Yu Li}
\affiliation{Yunnan Observatories, Chinese Academy of Sciences, Kunming 650216, People's Republic of China}

\author[0000-0002-5995-0794]{Sheng-Bang Qian}
\altaffiliation{E-mail: qiansb@ynu.edu.cn}
\affiliation{Department of Astronomy, School of Physics and Astronomy, Yunnan University, Kunming 650091, People's Republic of China}
\affiliation{Key Laboratory of Astroparticle Physics of Yunnan Province, Yunnan University, Kunming 650091, People's Republic of China}

\author{Ai-Ying Zhou}
\affiliation{National Astronomical Observatories, Chinese Academy of Sciences, A20 Datun Road, Chaoyang District, Beijing 100101, People's Republic of China}

\author[0000-0002-0796-7009]{Li-Ying Zhu}
\affiliation{Yunnan Observatories, Chinese Academy of Sciences, Kunming 650216, People's Republic of China}
\affiliation{University of Chinese Academy of Sciences, No.1 Yanqihu East Road, Huairou District, Beijing 101408, People's Republic of China}

\author[0000-0001-9346-9876]{Wen-Ping Liao}
\affiliation{Yunnan Observatories, Chinese Academy of Sciences, Kunming 650216, People's Republic of China}
\affiliation{University of Chinese Academy of Sciences, No.1 Yanqihu East Road, Huairou District, Beijing 101408, People's Republic of China}

\author{Er-Gang Zhao}
\affiliation{Yunnan Observatories, Chinese Academy of Sciences, Kunming 650216, People's Republic of China}

\author[0000-0002-5038-5952]{Xiang-Dong Shi}
\affiliation{Yunnan Observatories, Chinese Academy of Sciences, Kunming 650216, People's Republic of China}

\author[0000-0002-0285-6051]{Fu-Xing Li}
\affiliation{Department of Astronomy, School of Physics and Astronomy, Yunnan University, Kunming 650091, People's Republic of China}
\affiliation{Key Laboratory of Astroparticle Physics of Yunnan Province, Yunnan University, Kunming 650091, People's Republic of China}

\author[0000-0003-0516-404X]{Qi-Bin Sun}
\affiliation{Department of Astronomy, School of Physics and Astronomy, Yunnan University, Kunming 650091, People's Republic of China}
\affiliation{Key Laboratory of Astroparticle Physics of Yunnan Province, Yunnan University, Kunming 650091, People's Republic of China}



\begin{abstract}

Heartbeat stars (HBSs) with tidally excited oscillations (TEOs) are ideal astrophysical laboratories for studying the internal properties of the systems. In this paper, five new HBSs exhibiting TEOs are discovered using TESS photometric data. The orbital parameters are derived using a corrected version of Kumar et al.'s model based on the Markov Chain Monte Carlo (MCMC) method. The TEOs in these objects are examined, and their pulsation phases and modes are identified. The pulsation phases of the TEOs in TIC 266809405, TIC 266894805, and TIC 412881444 are consistent with the dominant $l=2$, $m=0$, or $\pm2$ spherical harmonic. For TIC 11619404, although the TEO phase is close to the $m=+2$ mode, the $m = 0$ mode cannot be excluded because of the low inclination in this system. The TEO phase in TIC 447927324 shows a large deviation ($>2\sigma$) from the adiabatic expectations, suggesting that it is expected to be a traveling wave rather than a standing wave. In addition, these TEOs occur at relatively low orbital harmonics, and we cautiously suggest that this may be an observational bias. These objects are valuable sources for studying the structure and evolution of eccentricity orbit binaries and extending the TESS HBS catalog with TEOs.

\end{abstract}



\section{Introduction} \label{sec:intro}
Heartbeat stars (HBSs) are a subclass of detached binaries with eccentric orbits. They are named for the shape of their light curve, which resembles an electrocardiogram \citep{2012ApJ...753...86T}. Meanwhile, tidally excited oscillations (TEOs) are an essential property of HBSs that can be used to probe the internal structure of the star \citep{2020ApJ...888...95G}. They are induced by the phase-dependent tides in binary stars with eccentric orbits \citep{1975A&A....41..329Z,1995ApJ...449..294K,2017MNRAS.472.1538F,2021FrASS...7..102J,2022A&A...659A..47K}, and can occur in some of the HBSs. Theoretical work is extensive, but it is challenging to detect with ground-based telescopes due to the small amplitude and short duration variations in their light curves \citep{2017MNRAS.472.1538F}. Since the Kepler satellite \citep{2010Sci...327..977B} provides long baselines and high-precision photometric data for such small variations, several Kepler HBSs and their TEOs have been studied in a series of papers \citep{2011ApJS..197....4W, 2012MNRAS.420.3126F, 2012MNRAS.421..983B, 2014MNRAS.440.3036O, 2013MNRAS.434..925H, 2016MNRAS.463.1199H, 2017ApJ...834...59G, 2019ApJ...885...46G, 2020ApJ...888...95G, 2020ApJ...896..161G, 2022MNRAS.517..437G, 2024MNRAS.530..586L, 2024ApJ...962...44L}.

On the other hand, the Transiting Exoplanet Survey Satellite (TESS; \cite{2015JATIS...1a4003R}) also provides a valuable opportunity to study HBSs and their TEOs. \citet{2021A&A...647A..12K} studied 20 TESS HBSs, seven of which have TEOs. \citet{2023AJ....166...42W} reported the linear and nonlinear TEOs and $\delta$ Sct pulsations in HBS FX UMa. The interesting HBS MACHO 80.7443.1718 (TIC 373840312) has also been studied in a number of papers \citep{2019MNRAS.489.4705J, 2021MNRAS.506.4083J, 2023A&A...671A..22K, 2023NatAs.tmp..173M, 2024A&A...686A.199K}. However, given the high resolution and continuous release of TESS photometric data, the potential of studying HBSs and their TEOs based on TESS data has not yet been fully exploited.

Thanks to the continuous release of the TESS data, we have discovered five new TESS HBSs and their TEOs in this paper. Section \ref{sec:analysis} presents the analytic procedure for these objects, including the HBS modeling, the TEO detection approach, and the pulsation phases and mode identification of the TEOs. Section \ref{sec:rst} presents the results of the analytic procedure. Section \ref{sec:discussion} discusses the results and concludes our work.

\section{Data and Analyses} \label{sec:analysis}
\subsection{Data Reduction}
The TESS was launched in 2018 as a space-based, all-sky optical exoplanet survey mission. The field of view is 24 $\times$ 96 degrees, and the observation sector is approximately 27 days for a given position in the sky. We use the light curve data processed by TESS-SPOC (the TESS Science Processing Operations Center) and the MIT QLP (the MIT Quick-Look Pipeline), downloaded by using the {\tt\string lightkurve} package \citep{2018ascl.soft12013L}. We visually inspected the light curves of each object and selected data sources with low scatter and significant heartbeat signals for analysis. Columns 5 and 6 in Table \ref{tab:basicparms} represent the selected data source and sector(s). We also removed obvious outliers by visual inspection and detrended the data using the Locally Weighted Scatter-plot Smoothing (LOWESS) approach \citep{cleveland1979robust}.

\subsection{Modeling}\label{sec:modeling}
We fit the corrected version of the \citet{1995ApJ...449..294K} model (K95$^+$ model) to the light curves. The K95$^+$ model is shown in Eq. (\ref{equation:one}) and differs from Eq. (44) in \citet{1995ApJ...449..294K} in that the sign before $\omega$ is changed from minus to plus \citep{2022ApJ...928..135W}:
\begin{equation}\label{equation:one}
	\frac{\delta F}{F}(t) = S\cdot\frac{1-3\sin^2i\sin^2(\varphi(t)+\omega)}{(R(t)/a)^3}+C,
\end{equation}
The K95$^+$ model approximates the equilibrium tidal deformation with a sum of all dominant modes, with $l=2$ and $m = 0, \pm2$ \citep{2021A&A...647A..12K}, and contains seven parameters: orbital period ($P$), eccentricity ($e$), orbital inclination ($i$), argument of periastron ($\omega$), the epoch of periastron message ($T_{0p}$), the amplitude scaling factor ($S$), and the fractional flux offset ($C$) \citep{2023ApJS..266...28L}.

We use the Markov Chain Monte Carlo (MCMC) method with the {\tt\string emcee} v3.1.2 Python package \citep{2013PASP..125..306F} to fit the light curves, following the fitting approach in \citet{2023ApJS..266...28L}. The parameters and their uncertainties of the five HBSs are derived and presented in Table \ref{tab:hbparms}, where the first column is the TESS ID; columns 2$-$8 represent the seven parameters of the K95$^+$ model.

\subsection{Detection of TEOs}\label{sec:teos}
We then use the analytic procedure in \citet{2024ApJ...962...44L} to examine the harmonic TEOs in these HBSs. The Fourier spectra are derived using the FNPEAKS \footnote{\url{http://helas.astro.uni.wroc.pl/deliverables.php?active=fnpeaks}} code. The mean noise level of each frequency, N, is derived as the mean amplitude in the frequency range $\pm$1 d$^{-1}$. Only frequencies with a signal-to-noise ratio (S/N) greater than 4.0 are used for TEO analysis. If the frequency $f$ satisfies at least one of the following equations, it is considered a harmonic TEO candidate:
\begin{equation}\label{equation:a}
	|n-f/f_{\rm orb}| < 0.01,
\end{equation}
\begin{equation}\label{equation:b}
	|n-f/f_{\rm orb}| < 3\sigma_{f/f_{\rm orb}}, 
\end{equation}
where $n$ is the harmonic number, $f_{\rm orb}=1/P$ is the orbital frequency, $\sigma_{f/f_{\rm orb}}=\sqrt{P^2\sigma_f^2+f^2\sigma_P^2}$ is the uncertainty of $f/f_{\rm orb}$ according to \citet{2021A&A...647A..12K} and \citet{2022ApJ...928..135W}, $\sigma_f$ and $\sigma_P$ stand for the uncertainties of $f$ and $P$, respectively. $\sigma_f$ is estimated following \citet{2008A&A...481..571K}. $P$ and $\sigma_P$ are represented in Table \ref{tab:hbparms}. Following the analytic procedure \citep{2024ApJ...962...44L}, we derive the harmonic TEO for each HBS. Since there is only one harmonic TEO for each HBS, we combine them into Table \ref{tab:TEOs_Phase} for presentation.

\subsection{Pulsation phases and mode identification}\label{sec:phases}
We then further identify the pulsation phases and mode of these TEOs following \citet{2024MNRAS.530..586L}. The pulsation phases of the TEOs for dominant modes of spherical harmonic degree $l=2$ can be expressed by Equation (\ref{equation:phi}) is based on the following assumptions: (1) the axes of the pulsation, spin, and orbit are all aligned; (2) the pulsations are adiabatic and the TEOs are standing waves; (3) the TEOs are not fine-tuned \citep{2014MNRAS.440.3036O, 2020ApJ...888...95G}:
\begin{equation} \label{equation:phi}
	\phi_{_{l=2,m}}=0.25+m\phi_{_{0}},
\end{equation}
where azimuthal order $m = 0, \pm2$, $\phi_{_{0}}=0.25-\omega/360^{\circ}$; $\omega$ is the argument of periastron. All phases can be measured with respect to the epoch of periastron passage $T_{0p}$ and are in units of 360$^{\circ}$. In addition, $\omega$ and $T_{0p}$ are presented in Table \ref{tab:hbparms}. We also perform a standard prewhitening procedure using Period04 \citep{2005CoAst.146...53L}, which calculates the uncertainties according to \citet{1999DSSN...13...28M}, to derive the phases of the harmonic frequencies. Note that since the phase is measured relative to $T_{0p}$ in the Fourier spectrum, the time of each data point in the photometric light curves should subtract $T_{0p}$ before Fourier analysis. Following the analytic procedure in \citet{2024MNRAS.530..586L}, we identify the pulsation phases and mode of these TEOs and show them in each panel (d) of Figures \ref{fig:11619404} $-$ \ref{fig:447927324}.

\section{Results} \label{sec:rst}
Table \ref{tab:basicparms} shows the basic parameters of the five TESS HSBSs, where columns 5 and 6 indicate the data source and sectors used in this paper.

\startlongtable
\begin{deluxetable*}{rccccc}
	\tablecaption{Basic parameters of the five TESS HBSs.
		\label{tab:basicparms}}
	\tablehead{
		\colhead{TESS ID} & \colhead{Simbad main ID} & \colhead{Coords (J2000)} & \colhead{V(mag)}  & \colhead{D.S.} & \colhead{sector(s)}
	}
	\startdata
	11619404 & HD 188060B & 19 52 19.62 +25 51 43.59 & 8.45 & (2) & 41 \\ 
	266809405 & HD 62629 & 07 45 08.94 +02 27 34.91 & 9.19 &  (1) & 34 \\ 
	266894805 & HD 62811 & 07 45 47.63 +02 33 35.46 & 7.97 &  (1) & 34 \\ 
	412881444 & BD+63 1793 & 21 58 43.06 +64 37 55.89 & 9.41 & (1) & 57,58 \\ 
	447927324 & HD 77812 & 09 03 07.87 $-$45 22 24.43 & 9.03 & (1) & 62 \\
	\enddata
	\tablecomments{The first column is the TESS ID; columns 2$-$4 are the basic parameters of the objects. Column 5 shows the data source: (1) TESS-SPOC data; (2) MIT QLP data. Column 6 shows the sector(s) used for analysis.}
\end{deluxetable*}

Table \ref{tab:hbparms} shows the parameters of the K95$^+$ model derived in section \ref{sec:modeling}.

\startlongtable
\begin{deluxetable*}{rrrrrrrr}
	\tablecaption{Parameters of the K95$^+$ model fitted to the light curves of the five TESS HBSs.
		\label{tab:hbparms}}
	\tablehead{
		\colhead{TESS ID} & \colhead{$P$(d)} & \colhead{$e$} & \colhead{$i$($^\circ$)} & \colhead{$\omega$($^\circ$)}  & \colhead{$T_{0p}$(TJD)} & \colhead{$S$($\times$10$^{-4}$)} & \colhead{$C$($\times$10$^{-4}$)} 
	}
	\startdata
	11619404 & 5.41884(57) & 0.3650(29) & 11.99(38) & 176.1(11) & 2423.1719(19) & 36.28(58) & -26.24(36) \\ 
	266809405 & 9.9644(51) & 0.4735(20) & 46.11(23) & 24.8(10) & 2235.555(10) & 4.439(50) & -0.573(62) \\ 
	266894805 & 5.78914(57) & 0.39915(81) & 50.41(11) & 66.11(22) & 2234.0678(19) & 8.319(43) & -0.771(41)  \\ 
	412881444 & 11.26990(71) & 0.46013(77) & 53.31(11) & 134.54(25) & 2856.9588(31) & 7.300(36) & -0.203(48)  \\ 
	447927324 & 3.26596(40) & 0.1865(14) & 29.574(96) & 154.24(41) & 2989.7078(29) & 29.44(21) & -17.05(18)  \\ 
	\enddata
	\tablecomments{The first column is the TESS ID; columns 2$-$8 are the seven parameters in the K95$^+$ model. The unit of $T_{0p}$ is TJD=BJD$-$2,457,000.}
\end{deluxetable*}

Table \ref{tab:TEOs_Phase} shows the TEO parameters of these HBSs, where columns 2 $-$ 4 are derived from section \ref{sec:teos}, and columns 5 $-$ 8 are derived from section \ref{sec:phases}.

\startlongtable
\begin{deluxetable*}{lrrrrrrr}
	\tabletypesize{\scriptsize}
	\label{tab:TEOs_Phase}
	\tablecaption{TEO parameters of the five TESS HBSs.}
	\tablehead{
		\colhead{TESS ID}& \colhead{$n$} & \colhead{$\Delta n$} &  \colhead{$3\sigma-|\Delta n|$} & \colhead{Frequency} & \colhead{Amplitude}  & \colhead{Phase} & \colhead{$S/N$} \\
		{} & {} & {} &{} & \colhead{(day$^{-1}$)} & \colhead{(mmag)} &{} 
	} 
	\startdata
	11619404 & 6 & -0.001 & 0.035355 & 1.10758(75) & 0.545(17) & 0.808(5) & 6.51 \\ 
	266809405 & 6 & 0.047 & 0.068229 & 0.6052(11) & 0.316(12) & 0.912(6) & 4.81 \\ 
	266894805 & 9 & 0.009 & 0.034922 & 1.5556(10) & 0.2316(85) & 0.129(6) & 6.73 \\ 
	412881444 & 12 & 0.004 & 0.028905 & 1.06476(33) & 0.1968(62) & 0.215(5) & 6.63 \\ 
	447927324 & 3 & 0.010 & 0.016348 & 0.92148(57) & 0.3564(90) & 0.455(4) & 5.1 \\ 
	\enddata
	\tablecomments{The first column is the TESS ID; $n$ is the harmonic number of the TEO; $\Delta n=f/f_{\rm orb}-n$, where $f$ is the detected frequency (column 5); a positive value in column 4 indicates that it satisfies Eq. (\ref{equation:b}); column 6 is the amplitude; column 7 is the phase; $S/N$ is the signal-to-noise ratio.}
\end{deluxetable*}

\begin{figure}
	\centering
	\includegraphics[width=0.5\columnwidth]{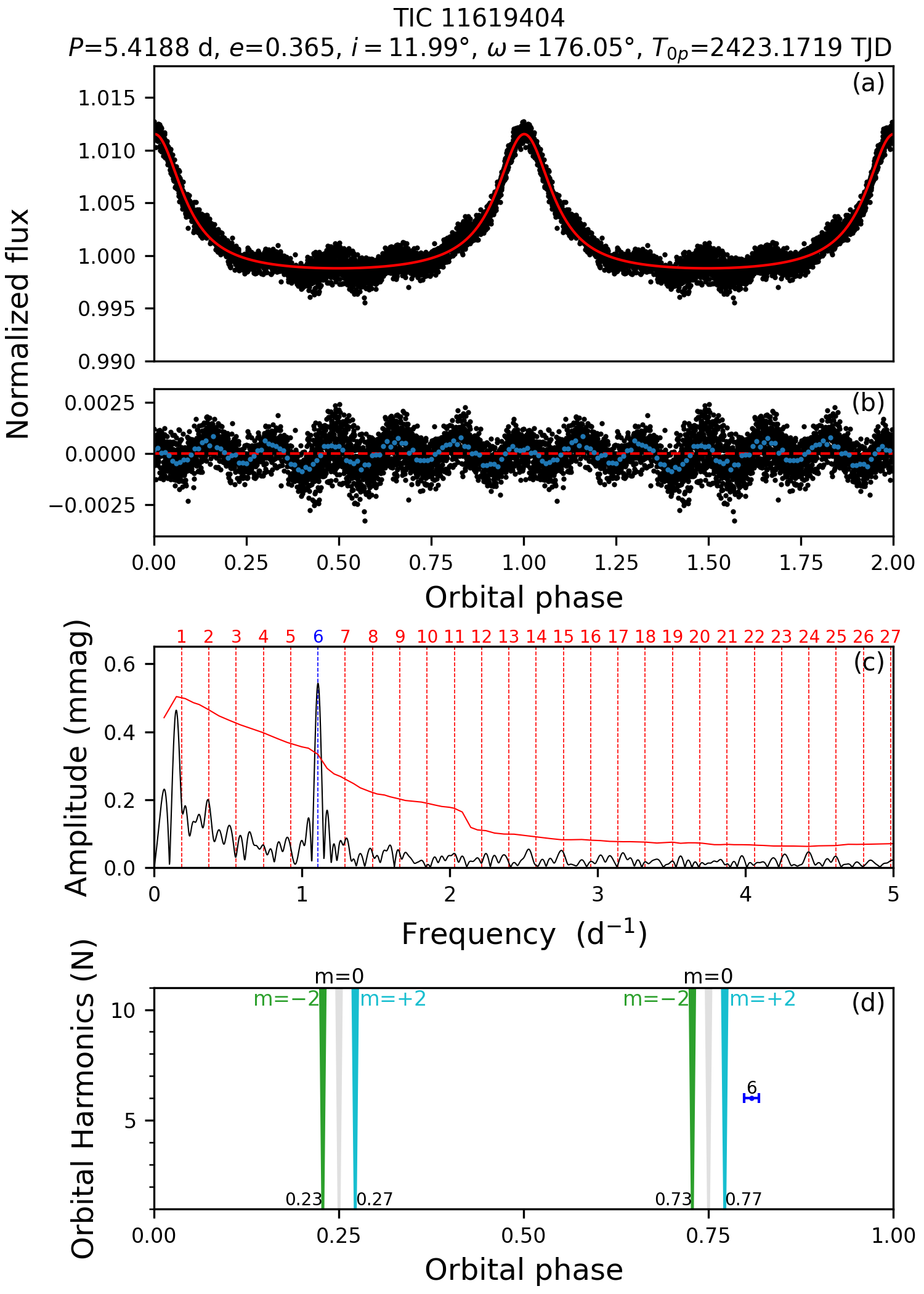}
	\caption{The analytic procedure for TIC 11619404. Panel (a): The K95$^+$ model (solid red line) fitted to the phase-folded light curve (black dots). Panel (b): The residuals of the fit in panel (a), and the blue dots are medians in 0.01 phase bins. Panel (c): The Fourier spectrum of the residuals from panel (b). The red and blue vertical dashed lines represent the orbital harmonics n; the blue lines indicate that they are harmonic TEOs. The solid red line shows the level of S/N = 4.0. Panel (d): The pulsation phases of the TEOs. The gray, light blue, and green strips indicate the $m=0,+2$, and $-2$ modes, respectively. The phases of the $m=+2,-2$ modes are shown next to the strips. The width of the strips results from the uncertainties of $T_{0p}$ and $\omega$. The blue circle represents a TEO with its harmonic number $n$; the size corresponds to its amplitude; the error bar corresponds to the uncertainty of its phase.
		\label{fig:11619404}}
\end{figure}

\subsection{TIC 11619404 (Figure \ref{fig:11619404})}
This is a 5.4 day HBS system, with an eccentricity of 0.365 and a low inclination of 11$^{\circ}$.99. The TEOs are obvious in panel (b), and panel (c) shows that the harmonic number is $n$ = 6. Since the $\omega$ of 176$^\circ$.1 is very close to 180$^\circ$, the $m=\pm2$ strips are close to the gray strips, making it difficult to identify the modes. As shown in panel (d), the $n$ = 6 harmonic is close to the $m=+2$ mode. However, \citet{2020ApJ...888...95G} have concluded that a low inclination (less than 30$^{\circ}$) would favor the $m = 0$ mode. Given the low inclination in this system, we suggest that the $m = 0$ mode cannot be excluded either.

\begin{figure}
	\centering
	\includegraphics[width=0.5\columnwidth]{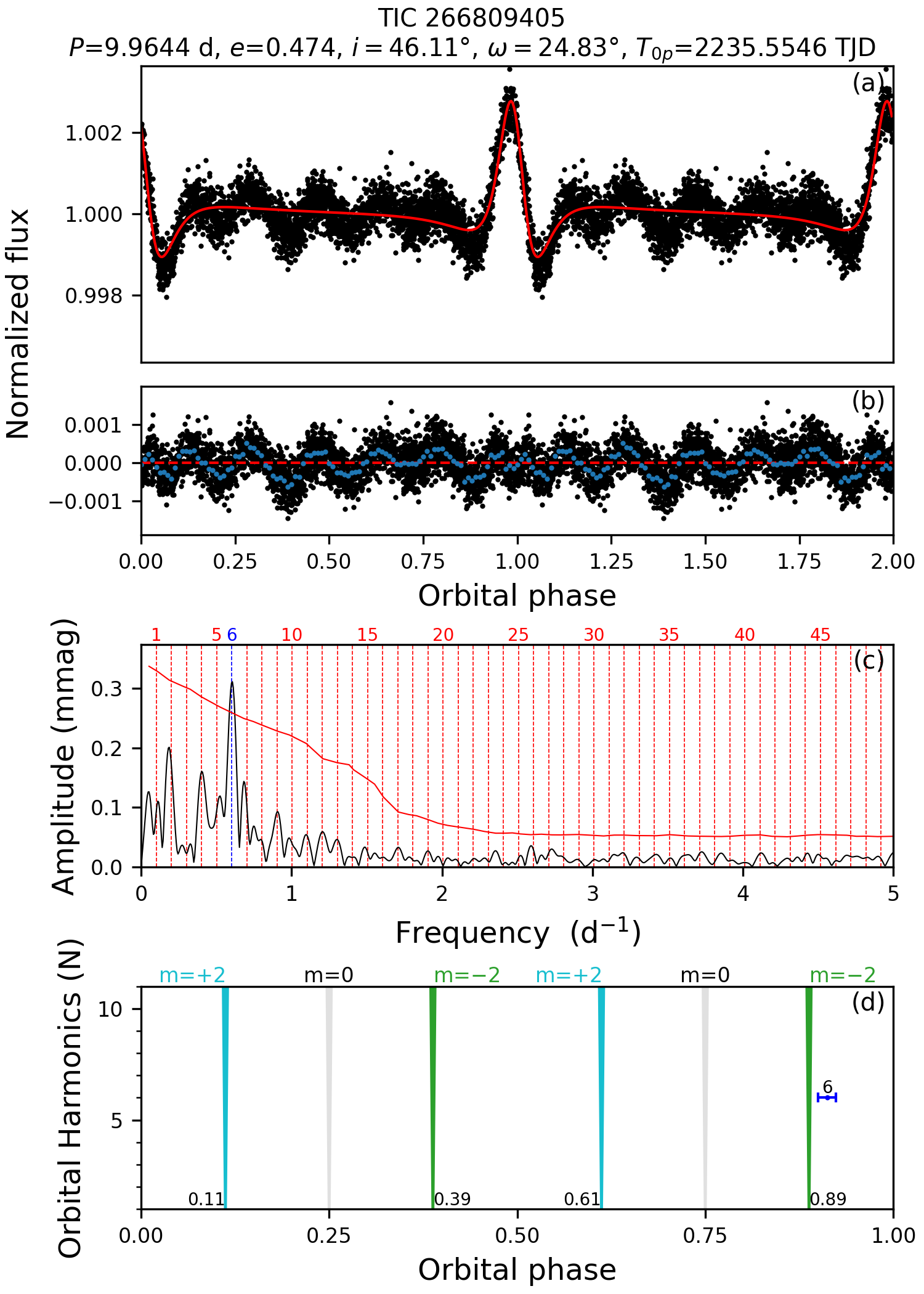}
	\caption{Same as Fig. \ref{fig:11619404} for TIC 266809405.
		\label{fig:266809405}}
\end{figure}

\subsection{TIC 266809405 (Figure \ref{fig:266809405})}
This is a period of 9.96 day HBS system with an eccentricity of 0.474, an intermediate inclination of 46$^{\circ}$.1, and an argument of periastron $\omega$ of 24$^\circ$.8. The $n$ = 6 orbital harmonic stands out clearly in panel (c) of Figure \ref{fig:266809405}. Panel (d) shows that the phase of the $n$ = 6 harmonic is consistent with the $m=-2$ mode.

\begin{figure}
	\centering
	\includegraphics[width=0.5\columnwidth]{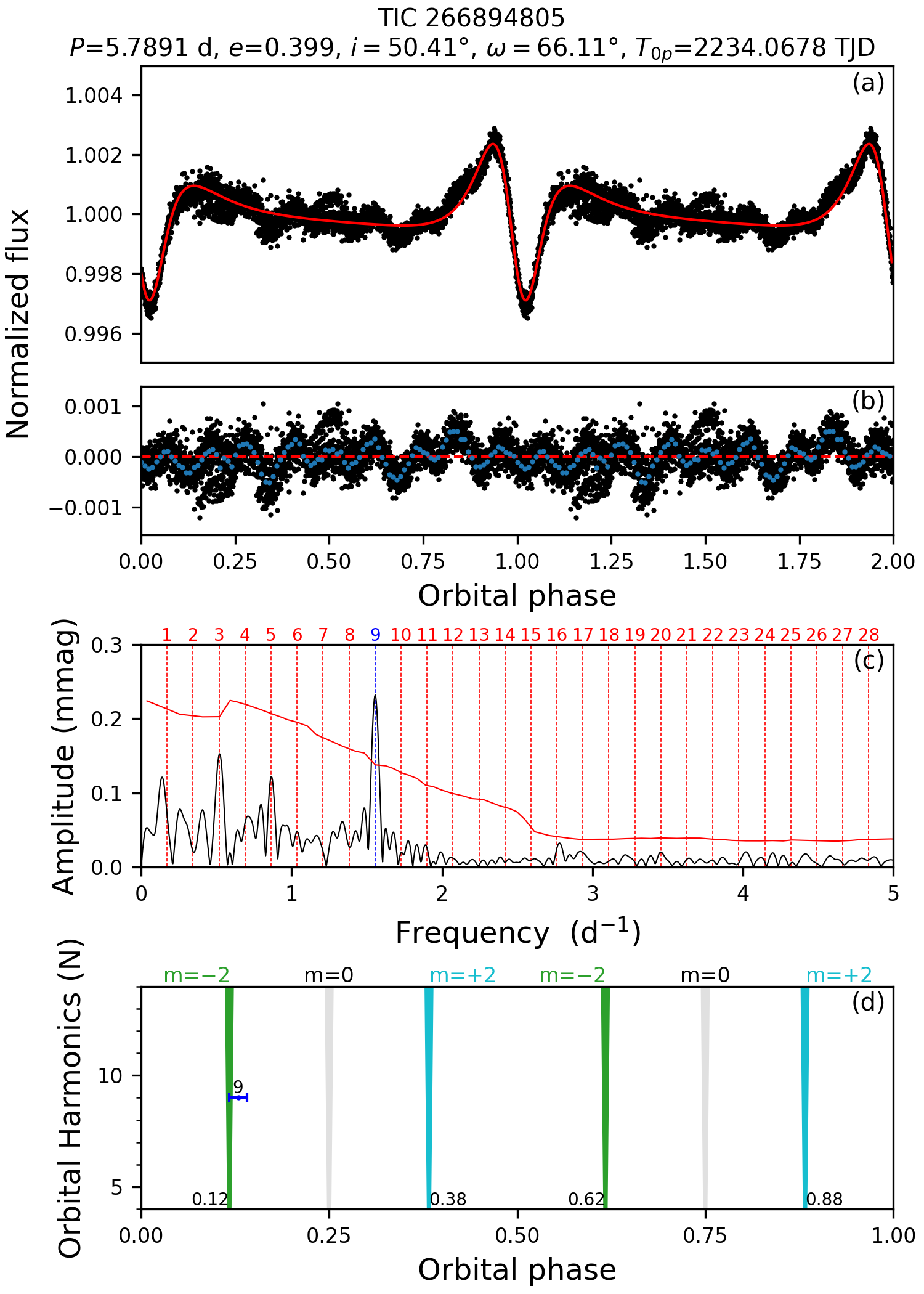}
	\caption{Same as Fig. \ref{fig:11619404} for TIC 266894805.
		\label{fig:266894805}}
\end{figure}

\subsection{TIC 266894805 (Figure \ref{fig:266894805})}
This is a 5.8 day HBS system, with an eccentricity of 0.399, an intermediate inclination of 50$^{\circ}$.41, and an argument of periastron $\omega$ of 66$^\circ$.11. The $n$ = 9 orbital harmonic stands out clearly in panel (c) of Figure \ref{fig:266894805}. Panel (d) shows that the phase of the $n$ = 9 harmonic is consistent with the $m=-2$ mode.

\begin{figure}
	\centering
	\includegraphics[width=0.5\columnwidth]{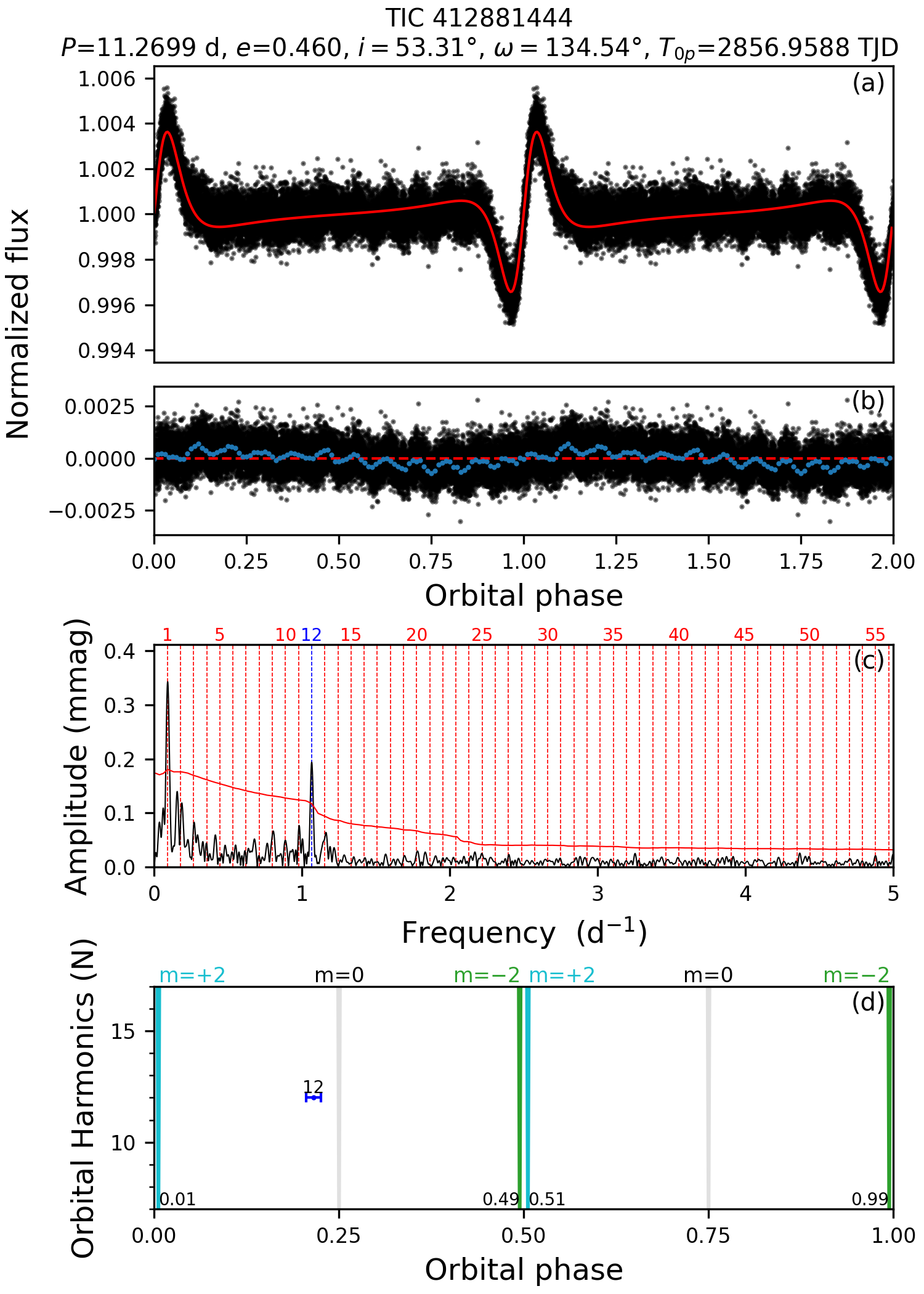}
	\caption{Same as Fig. \ref{fig:11619404} for TIC 412881444.
		\label{fig:412881444}}
\end{figure}

\subsection{TIC 412881444 (Figure \ref{fig:412881444})}
This is a period of 11.3 day HBS system, with an eccentricity of 0.460, an inclination of 53$^{\circ}$.31, and an argument of periastron $\omega$ of 134$^\circ$.54, as shown in Figure \ref{fig:412881444}. The $n$ = 12 harmonic is clearly visible in panel (c). Panel (d) shows that its pulsation phase is close to the $m = 0$ mode.

\begin{figure}
	\centering
	\includegraphics[width=0.5\columnwidth]{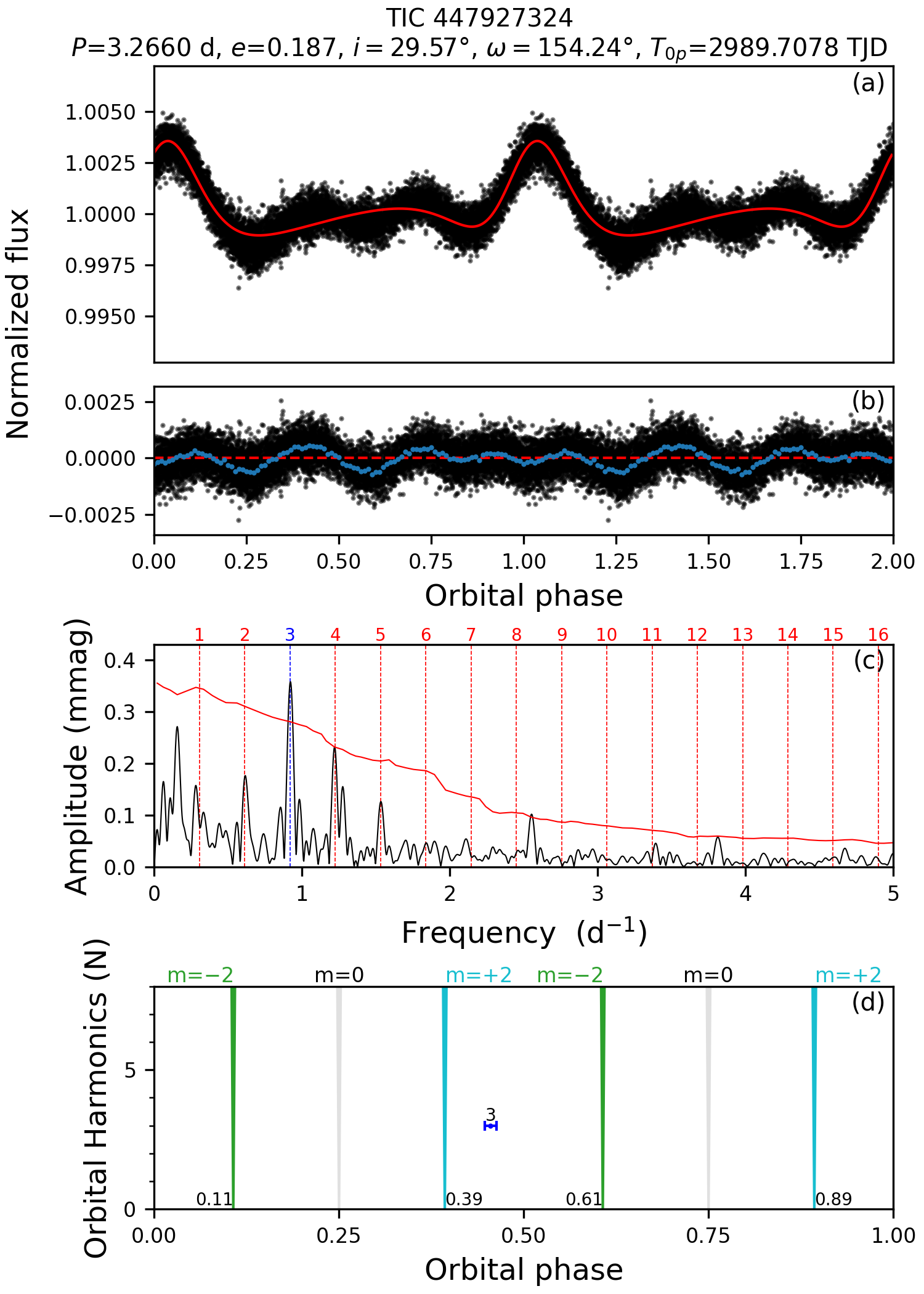}
	\caption{Same as Fig. \ref{fig:11619404} for TIC 447927324.
		\label{fig:447927324}}
\end{figure}

\subsection{TIC 447927324 (Figure \ref{fig:447927324})}
This is a short-period of 3.3 day HBS system, with a low eccentricity of 0.187, a low inclination of 29$^{\circ}$.57, and an argument of periastron $\omega$ of 154$^\circ$.24, as shown in Figure \ref{fig:447927324}. Panel (c) shows that the TEO candidate is the $n$ = 3 harmonic. In addition, the $n$ = 4 and 5 harmonics also exist, but given their lower S/N ($\le$ 4.0), we exclude them as the TEO candidates. However, the $n$ = 3 harmonic shows a large deviation ($>2\sigma$) from the adiabatic expectations in panel (d). Given the presence of obvious TEOs in panels (a) and (b), we suggest that the $n$ = 3 harmonic is expected to be a traveling wave rather than a standing wave.

\section{Discussions and Conclusions} \label{sec:discussion}
The components in HBS systems are distorted by the time-varying tidal potential of the companion star, and the response of the components can usually be divided into two parts: the equilibrium tide and the dynamical tide. The equilibrium tide contributes to the ``heartbeat" feature, while the dynamic tide is made up of TEOs that are visible in all phases of the orbit \citep{2017MNRAS.472.1538F}. To study TEOs, the contribution of the equilibrium tide must first be subtracted from the light curves. The remaining dynamical tide, which contains features of TEOs, can then be used to perform Fourier spectral analysis \citep{2020ApJ...888...95G}. Meanwhile, the K95$^+$ model can be used to model these non-eclipsing HBSs \citep{2023ApJS..266...28L}. 

Note that the K95$^+$ model does not take into account the irradiation/reflection effects or Doppler beaming \citep{2021A&A...647A..12K}. Therefore, the derived parameters may show some deviations. However, based on previous works, we believe that such deviations are limited. First, the orbital period $P$, the eccentricity $e$, and the argument of periastron $\omega$ derived from the K95$^+$ model are usually reliable \citep{2020ApJ...888...95G, 2023ApJS..266...28L}. Second, the inclination $i$ can indeed have deviations in different cases \citep{2022ApJ...928..135W, 2023ApJS..266...28L}. In particular, it is possible for $i$ to be overestimated if the obtained inclination is too high and there is no eclipse in the light curve. Third, $\omega$ derived from the different approaches may have some slight differences. In this case, a comprehensive consideration of these different values is required. Therefore, these differences do not significantly affect the mode identification results \citep{2024MNRAS.530..586L}. In summary, the results obtained from these approaches in this work are generally reliable.

In this work, we discover five new TESS HBSs that exhibit TEOs. First, we derive the orbital parameters using the K95$^+$ model based on the MCMC method. Second, we study the TEOs in these objects and then identify the pulsation phases and mode of the TEOs. The pulsation phases of the TEOs in TIC 266809405, TIC 266894805, and TIC 412881444 are consistent with the dominant $l=2$, $m=0$, or $\pm2$ spherical harmonic. For TIC 11619404, although the TEO phase is close to the $m=+2$ mode, the $m = 0$ mode cannot be excluded because of the low inclination in this system. The TEO phase in TIC 447927324 shows a large deviation ($>2\sigma$) from the adiabatic expectations, and we suggest that it is expected to be a traveling wave rather than a standing wave.

In addition, an interesting feature of these TEOs is that they occur at relatively low orbital harmonics. A similar situation was also found in the TESS HBSs reported by \citet{2021A&A...647A..12K}. For comparison, the median and maximum values of $n$ for TEOs in the TESS HBSs are lower than those in the Kepler HBSs reported by \citet{2024ApJ...962...44L}. We cautiously suggest that this may be an observational bias. After all, these TESS samples have relatively short continuous observing times, while the value for the Kepler HBSs is more than 1400 days. In our future work, we will aim to discover TESS HBSs with higher harmonic $n$ for TEOs.

\section*{Acknowledgements}

This work is partly supported by the International Cooperation Projects of the National Key R$\&$D Program (No. 2022YFE0127300), the National Natural Science Foundation of China (Nos. 11933008 and 12103084), the Basic Research Project of Yunnan Province (Grant Nos. 202201AT070092 and 202301AT070352), the Science Foundation of Yunnan Province (No. 202401AS070046), and the Yunnan Revitalization Talent Support Program. The NASA Explorer Program provides funding for the TESS mission. We thank the TESS teams for their support and hard work. We are grateful to the anonymous referee for reviewing this manuscript.

\software{
	{\tt\string emcee} \citep{2013PASP..125..306F},
	{\tt\string lightkurve} \citep{2018ascl.soft12013L},
	FNPEAKS (Z. Koo{\l}aczkowski, W. Hebisch, G. Kopacki),
	PERIOD04 \citep{2005CoAst.146...53L}.
}




\bibliography{sample631}
\bibliographystyle{aasjournal}




\end{document}